\documentstyle[psfig,mprocl]{article}
\begin{document}
\title{{\normalsize In: D. E. Wolf, P. Grassberger (Eds.) ``Friction,
Arching, Contact Dynamics'' World Scientific (Singapore, 1997) p. 265-273}\\[1.5cm]
Force Distribution and Comminution in Ball Mills}
\author{Volkhard Buchholtz and Thorsten P\"oschel}
\address{Institut f\"ur Physik, Humboldt-Universit\"at zu Berlin, 
Invalidenstr. 110, 10115 Berlin\\
email: volkhard\@summa.physik.hu-berlin.de, 
thorsten\@itp02.physik.hu-berlin.de}
\maketitle

\abstracts{ The motion of granular material in a ball mill is
  investigated using molecular dynamics simulations in two dimensions. In
  agreement with experimental observations by
  Rothkegel\,\cite{Rothkegel} we find that local stresses -- and hence
  the comminution efficiency -- are maximal close to the bottom of the
  container. This effect will be explained using analysis of
  statistics of force chains in the material.  }

\section{Introduction}
Milling is one of the most important techniques in mechanical
engineering and much effort has been done to optimize the comminution
efficiency of milling techniques with respect to energy, space and
time consumption\,\cite{AllgBuch}.

Mainly because of their simple construction and application ball
milling is a wide spread milling technology, particularly in mining.
Ball mills are (usually large) cylindrical devices of length and
diameter up to several meters which revolve along their axis. The
material to be comminuted moves inside the cylinder. There are two
different methods: autogeneous comminution where the material is pure
and heterogeneous comminution where the material is mixed with heavy
spheres from steel of typical diameter of several centimeters to
increase efficiency. Throughout the paper we will deal with autogeneous
comminution. Unfortunately the efficiency of ball mills is not very
high and, therefore, engineers did much scientific work to increase
their efficiency. There exists much experimental knowledge on the
operation mechanisms of ball mills and on the comminution of granular
matter in these mills. An overview can be found in
ref.\,\cite{AllgBuch}.

There are still effects, however, which have not been understood yet.
The present paper deals with the spatial distribution of stress in the
material. At first glance one could assume that the largest stresses
will be observed close to the surface of the material where the
granular particles have the largest values of relative collision
velocity. Hence one could assume that large part of particle
comminution occurs close to the material surface. To increase
efficiency of the mill one would have to tune the rotation velocity so
that the average collision velocity becomes maximum.

Experimental investigations, however, show opposite results: Rothkegel
and Rolf measured directly the spatial distribution of intensive
impacts\,\cite{Rothkegel} in a ball mill. In an experiment using a
small almost two-dimensional mill they used instrumented balls which
flash whenever the acting force at a particle contact is larger than a
threshold. The statistics of the spatial distribution of the flashes
led to a surprising result: the spatial distribution of intensive
particle contacts which in realistic ball mills might lead to particle
comminution has its maximum deep inside the material close to the
bottom of the cylinder. This result was new and there was no
satisfying explanation for the measured stress distribution yet.

In the following section we present the results of a two dimensional
molecular dynamics simulation of a ball mill. It will be shown that
the spatial stress distribution is closely related to the properties
of force chains in the material. The occurrence of such force chains,
and even of entire networks of such chains has been reported before,
see e.g. refs.\,\cite{FCDuran}$^-$\cite{FCEP}.

\section{Molecular dynamics model}
The rotation velocity of ball mills is high enough to keep the
granular particles in the continuous flow regime (for explanation of
the regimes see\,\cite{Rajchenbach}) and on top of the material grains
are thrown through the air caused by intensive forcing due to rapid
revolution of the mill. Therefore static friction does not play an
important role for the dynamics of the material and, hence, we can apply
a sphere model in the MD simulations. We just want to remark that
in other cases when static behavior of granular material becomes
essential for the dynamics one has to apply more complicated grain
models of non-spherical shape. For detailed discussions of the
differences in the dynamics of spherical and non-spherical objects see
references\,\cite{PBPRL}$^-$\cite{BPJSP}.

For our simulation we apply a spherical MD model for granular materials
by Cundall and Strack\,\cite{CS} and Haff and Werner\,\cite{HW} which
was used by many authors in simulations of rapidly moving granular
material.

Two particles $i$ and $j$ with radii $R_i$ and $R_j$, positions
$\vec{r}_i$, $\vec{r}_j$ and velocities $\dot{\vec{r}}_i$ and
$\dot{\vec{r}}_j$ feel a force 
\begin{equation}
  \label{force}
\vec{F}_{ij} = F_{ij}^N\, \vec{n} + F_{ij}^T\, \vec{t} 
\label{force1}
\end{equation}
only when contacting, i.e. if the condition $\left|
  \vec{r}_i-\vec{r}_j\right| < R_i+R_j$ holds. The force consist of a
component $F_{ij}^N$ acting in normal direction $\vec{n}$
\begin{equation}
  F_{ij}^N = Y \cdot \left(R_i+R_j- \left| \vec{r}_i-\vec{r}_j\right| 
  \right) - m_{ij}^{\mbox{\footnotesize \em eff}} \cdot \gamma_N \cdot 
  (\dot{\vec{r}}_i - \dot{\vec{r}}_j) \cdot\vec{n}
\label{fnormal}
\end{equation}

and a component $F_{ij}^T$ acting in tangential direction $\vec{t} =
\left( {0 ~ -1}\atop{1 ~~~~ 0} \right) \vec{n}$
 \begin{eqnarray}
F_{ij}^T &=& \mbox{sign} \left(v_{ij}^{\mbox{\footnotesize \em rel}}\right) 
\cdot \min\left( m_{ij}^{\mbox{\footnotesize \em eff}} \gamma_T \left|
v_{ij}^{\mbox{\footnotesize \em rel}}\right| , 
\mu \left| F_{ij}^{N}\right| \right) 
\label{ftang} \\
\mbox{with}&&\\
v_{ij}^{\mbox{\footnotesize \em rel}} &=& (\dot{\vec{r}}_i - 
\dot{\vec{r}}_j)  \cdot
\vec{t}+ R_i \cdot \omega_i + R_j \cdot \omega_j 
\label{surfvelocity}\\ 
m_{ij}^{\mbox{\footnotesize \em eff}} &=& \frac{m_i \cdot m_j}{m_i + m_j} ~.
\label{meff}
\end{eqnarray}

$m_i$ and $\omega_i$ are the mass and the rotation velocity of the
$i$-th particle. $Y=8\cdot 10^6~g~s^{-2}$ is the Young Modulus,
$\gamma_N=800~s^{-1}$ and $\gamma_T=3000~s^{-1}$ are the damping
coefficients in normal and tangential direction, and $\mu=0.5$ is the
Coulomb friction coefficient. The parameters have been chosen to give
realistic results in comparison of the simulation results with the
behavior of a typical granular material.

Eq.~(\ref{surfvelocity}) describes the relative velocity of the
particle surfaces at the contact point. This velocity consists of the
tangential part of the relative velocity of the particles and of a
term which originates from rotation of the grains.
$m_{ij}^{\mbox{\footnotesize \em rel}}$ stands for the effective mass.

\centerline{\psfig{figure=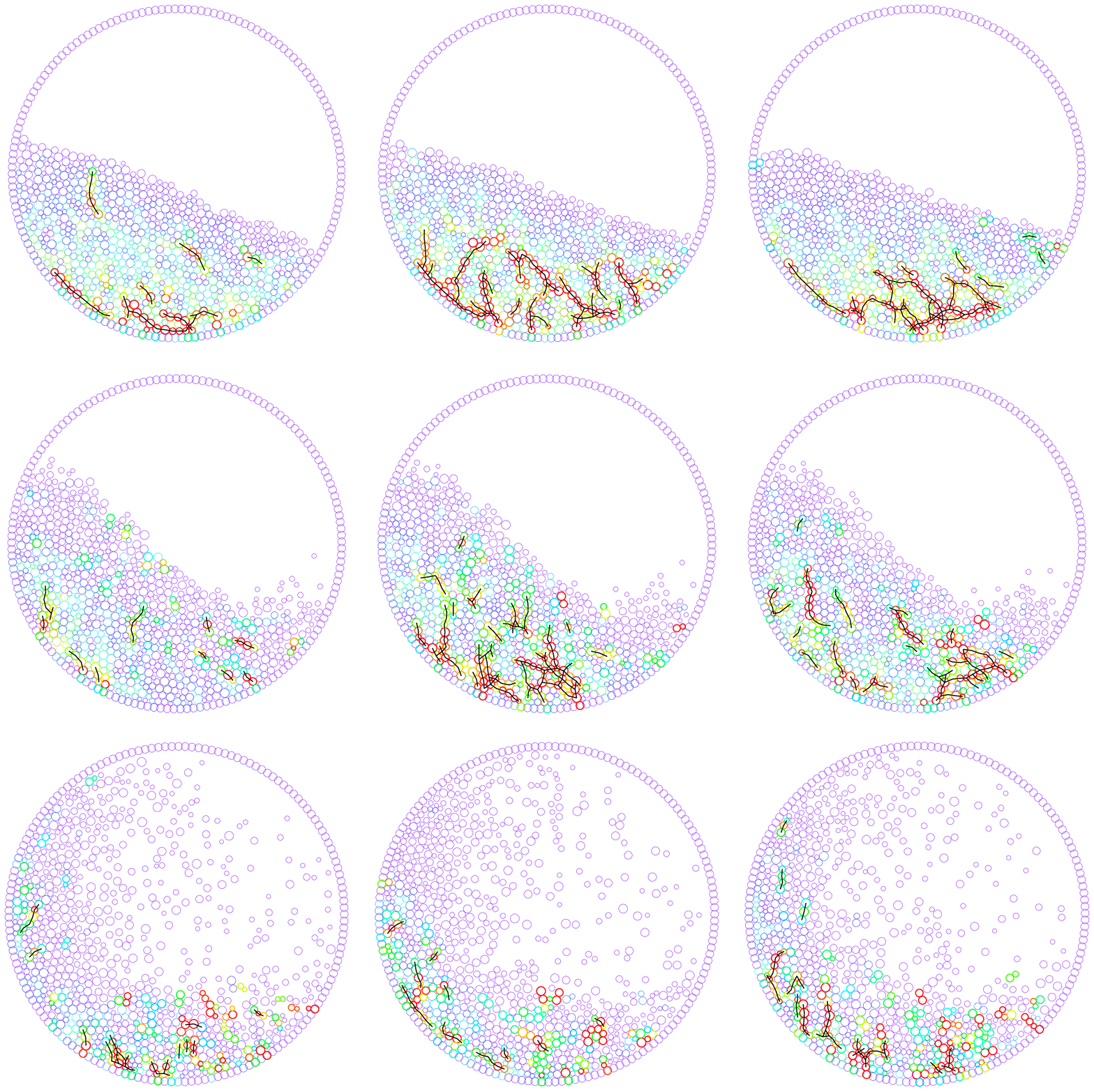,width=12.5cm}}
\begin{figure}[h]
  \caption{Snapshots of the simulation for different velocity. The 
    color codes for the local pressure $P_i$
    (cf.~eq.~(\ref{pressure})). Black lines connecting adjacent
    particles mark automatically detected force chains. The ball mill
    revolves with angular velocity $\Omega_{I} = 2 Hz$ (top),
    $\Omega_{II} = 10 Hz$ (middle), $\Omega_{III} = 19 Hz$ (bottom).}
  \label{fig:snapnew}
\end{figure}

The Coulomb friction law is taken into account by
equation~(\ref{ftang}) saying that two grains slide on each other if
the tangential force is larger than $\mu$ times the normal force.
Otherwise they roll mainly. For the detailed discussion of this and
other models for molecular dynamics simulation of granular material
see\,\cite{WolfForce}.

Based on the force (\ref{force1}) Newton's equations of motion have
been integrated using a Gear predictor-corrector scheme of fifth
order, e.g.\,\cite{AllenTildesley}. This method was used in many
simulations of granular material and has been proven to be very
numerically stable.

\section{Results of the simulation}

The model described in the previous section was used to simulate a
two-dimensional ball mill of diameter $D=8~cm$ filled with $N=800$
spherical grains. The radii of the grains are distributed in the
interval $[0.05,\,0.11]\,cm$. To study the influence of the rotation
velocity of the cylinder we performed simulations for three different
velocities $\Omega_{I} = 2 Hz = 2/(2\cdot\pi) \mbox{\,revolutions per
  second}$, $\Omega_{II} = 10 Hz = 10/(2\cdot\pi) \mbox{revolutions
  per second}$, $\Omega_{III} = 19 Hz = 19/(2\cdot\pi)
\mbox{revolutions per second}$.

At $\Omega=\Omega_{I}=2 Hz$ the flow at the surface is already
continuous and the surface is characterised by an inclined line (see
fig.~\ref{fig:snapnew}, top). For $\Omega=\Omega_{II}=10 Hz$ the material
surface is still almost compact and consists of two sections a steep
and a flat one (fig.~\ref{fig:snapnew}, middle). This velocity regime is
experimentally known to have the best comminution
efficiency\,\cite{AllgBuch}. Large rotation velocity
$\Omega=\Omega_{III}= 19 Hz$ leads to free flight of the grains at the
surface (fig.~\ref{fig:snapnew}, top).

Fig.~\ref{fig:snapnew} shows snapshots of simulations with $N=800$ grains for
the three mentioned rotation velocities $\Omega_{I}$, $\Omega_{II}$,
$\Omega_{III}$. The color codes for the local pressure $P_i$ acting on
the $i$-th particle
\begin{equation}
P_i = \sum\limits_j F_{ij}^N~.
\label{pressure}
\end{equation} 
The index $j$ runs over all neighbors of the $i$th particle.  Grains
feeling a pressure $P_i > 1000 g~cm~s^{-2}$ are drawn bold lined.
Obviously most of the hardly stressed particles are located near the
walls inside the bulk of material, which agrees well with the
experimental findings by Rothkegel and Rolf\,\cite{Rothkegel}.

The hypothesis we want to discuss in the following is, that the self
organized formation of {\em force chains} is responsible for the
spatial stress distribution and, hence, the relevant physical reason
for comminution processes in ball mills. Without the existence of such
force chains, ball mills would not work or at least would work much
less efficient. In the following we will discuss the properties of
these force chains.

Force chains are defined by a set of simple conditions: Particles are
called members of the same force chain if:
\begin{enumerate}
\item they feel a pressure (eq. (\ref{pressure})) of more than $1000\,
  g~cm~s^{-2}$,
\item each member particle of the chain touches at least one other
  member, and
\item for all members $k$ of the chain having two neighbors $i$ and
  $j$, which belong to the force chain as well, the condition
\begin{equation}
\left|\frac{\vec{r}_k-\vec{r}_i}{\left|\vec{r}_k-\vec{r}_i\right|}
  \cdot \frac{\vec{r}_k-\vec{r}_j}{\left|\vec{r}_k-\vec{r}_j\right|}
\right| < 0.85\,,
\end{equation}
holds, i.e. these three particles lie almost on a line.
\end{enumerate}

These three conditions can be checked using a computer algorithm. In
fig.~\ref{fig:snapnew} neighboring particles have been connected by a
black line provided that either of them belong to the same force
chain. Obviously most of the hard stressed (and therefore bold drawn)
particles are members of force chains. We conclude that an essential
part of the static and dynamic pressure in the ball mill propagates
along force chains. Particles which belong to a force chain feel
pressure which are up to 100 times larger than the local average
pressure
\begin{equation}
P_i^{\mbox{\footnotesize \em (av)}} = \frac{1}{K}\sum 
\limits_{k=1}^K P_k\,,
\end{equation}
where the index $k$ runs over all $K$ neighbors of the particle $i$
which are within a certain neighborhood
$\left|\vec{r}_i-\vec{r}_k\right| \le 4 \cdot \overline{R}$ and which
do {\em not} belong to any force chain.

The occurrence of force chains in granular systems is not new. They
have been observed before in other granular systems in
experiments\,\cite{FCDuran}$^-$\cite{FCLiu}
and computer
simulations\,\cite{FCWolf,FCEP}. It can be shown that one of the (at
least) three phases of an idealized granular system is characterized
by the occurrence of force chains\,\cite{FCEP}.

Fig.~\ref{hfvonl} (left) shows the frequency of force chains of
lengths $L$ where $L$ is the number of particles which join the chain.
The frequency decreases almost exponentially with increasing length.
For the highest rotation velocity $\Omega_{III}$ (fig.~\ref{hfvonl},
bottom) the frequency distribution breaks down for longer chains,
which is mostly due to less compact bulk of material and finite size
of the system.

\vspace{0.3cm}
\centerline{\psfig{figure=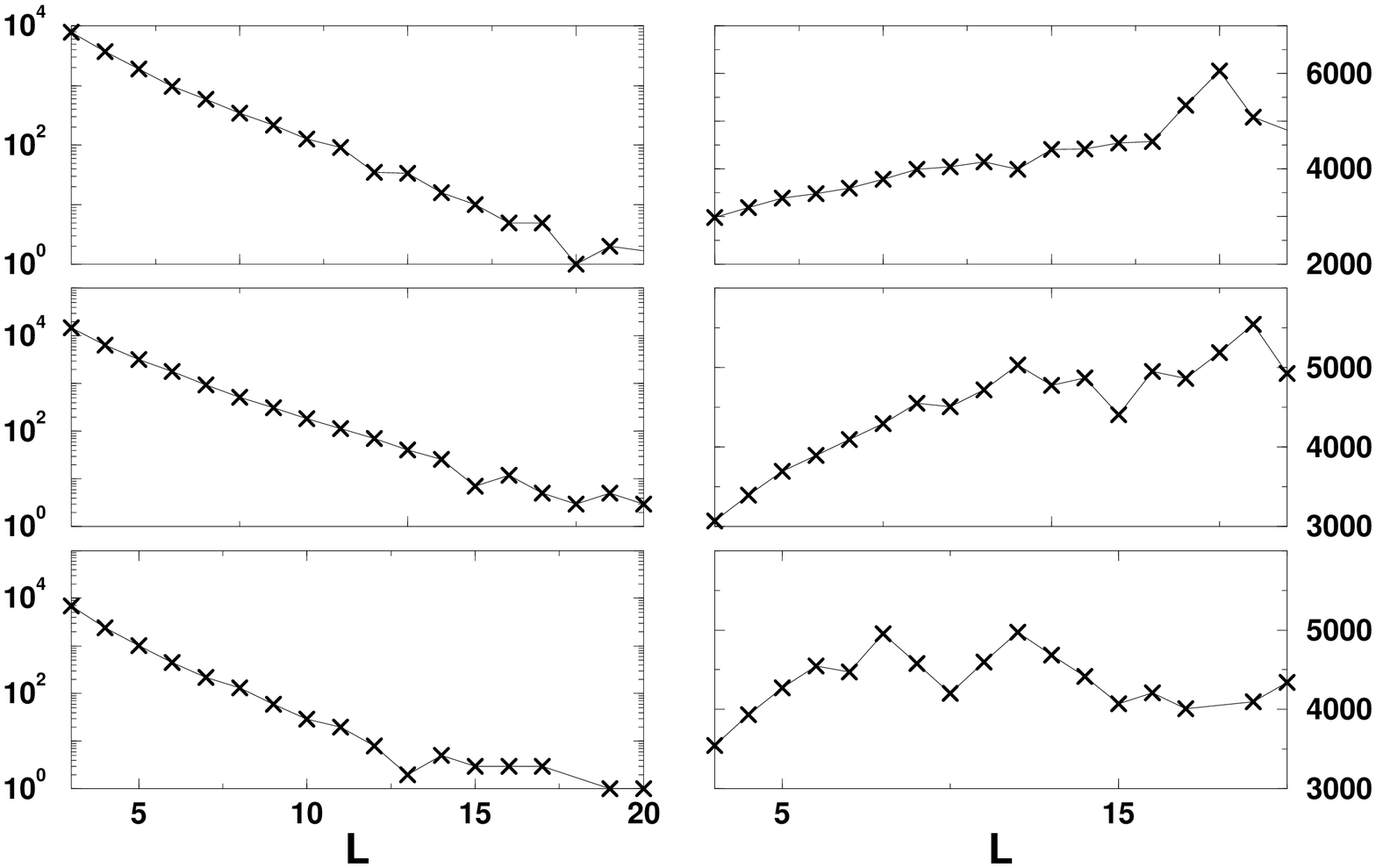,width=11cm,angle=270}}
\begin{figure}[h]
\vspace*{-0.5cm}
\caption{Left: Frequency of force chains of length $L$. Right: Average 
  of the maximal pressure inside one force chain as a function of the
  length $L$ of the force chain. Top: $\Omega_{I} = 2 Hz$, middle:
  $\Omega_{II} = 10 Hz$, bottom: $\Omega_{III} = 19 Hz$.}
\label{hfvonl}
\end{figure}

The right hand side of fig.~\ref{hfvonl} shows the correlation of the
length of a force chain and the pressure which acts on the particles
which belong to this chain. Of particular interest is not the {\em
  average} but the {\em maximum} pressure inside a chain. The maximum
pressure is criterion whether a grain will be comminuted in a mill. As
soon as one of the particles of a chain has broken the load discharges
and the force chain vanishes and reappears elsewhere. Therefore here
we are mainly interested in the {\em maximum} pressure. For low
velocities $\Omega_{I}$ and $\Omega_{II}$ (fig.~\ref{hfvonl}, top and
middle) the average value of the maximal force in a force chain of
length $L$ grows monotonously with the length, i.e. a force chain acts
similar as a beam in a framework: all the weight and momenta from
particles above the chain are supported by the chain and propagated
through the chain. Therefore one finds in most cases that the lower
the position of a grain the higher the load.

In contrast, the corresponding plot (fig.~\ref{hfvonl}, bottom) for
high velocity $\Omega_{III}$ shows no characteristic dependence of the
maximal force on the length of the force chain, i.e. for this case the
force chains do not act as described in the previous paragraph. Thus
we conclude that the mechanism of stress propagation in force chains
is much less efficient in the case of high rotation velocity.

In fig.~\ref{hfl} the average number of particles per force chain
which feel a pressure $P_i$ larger than a given threshold $P$ is
plotted versus the length of the force chain. This number is directly
related to the breaking probability which can be described using
Weibull statistics\,\cite{Weibull}. Again one can see that longer
force chains lead to a higher probability of comminution. For the
highest rotation velocity $\Omega_{III}$ the distributions are
significantly lower again.

\vspace{0.5cm}
\centerline{\psfig{figure=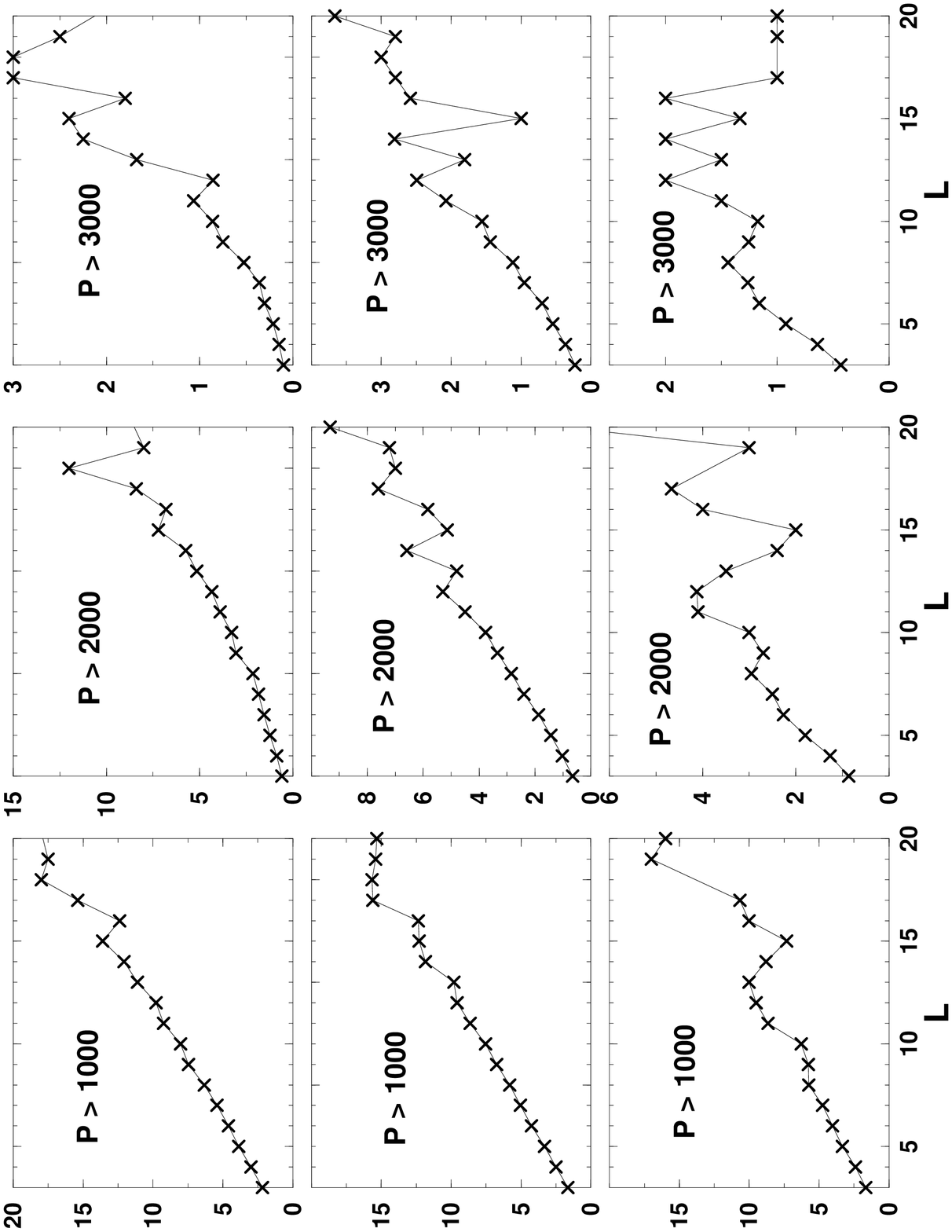,width=11cm,angle=270}}
\vspace{-0.2cm}
\begin{figure}[h]
\caption{Average number of grains per force chain, which feel a 
  local pressure $P_i$ larger than a given threshold $P$. Top:
  $\Omega_{I} = 2\, Hz$, middle: $\Omega_{II} = 10\, Hz$, bottom:
  $\Omega_{III} = 19\, Hz$. The longer the chain the more particles
  feel pressure $P_i>P$ and, hence, the higher the comminution
  probability.}
\label{hfl}
\end{figure}
\newpage

\begin{figure}[htbp]
\centerline{\psfig{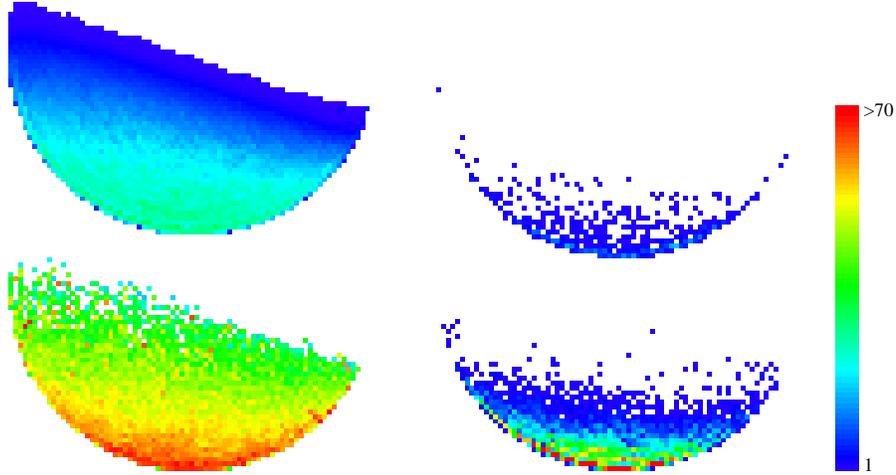}}
\caption{Left: Time averaged spatial distribution of the local 
  pressure for $\Omega_{I} = 2\, Hz$. The color codes for the pressure
  $P_i$, blue color means low pressure and red color stands for high
  pressure. The upper figure includes only particles which do not
  belong to a force chain, the lower figure includes only particles
  which do belong to a force chain. Right: Time averaged spatial
  distribution of the number of grains feeling a pressure $P_i >
  3000~cm~g~s^{-2}$. The color code for the right figures is given by
  the color bar. Again in the upper figure only grains which do not
  belong to a force chain have been considered, whereas in the lower
  figure only grains which belong to a force chain join the averaging
  procedure.}
\label{dist2}
\end{figure}

The time averaged spatial distributions of the pressure $P_i$ for
$\Omega_{I} = 2\, Hz$ are shown in fig.~\ref{dist2} (left part of
figure). The color codes for the local pressure $P_i$, where blue
means low pressure and red color stands for high pressure. The upper
plots include only particles which do {\em not} belong to a force
chain, whereas the lower figures include only particles which {\em do}
belong to a force chain.

The right hand side of figure~\ref{dist2} shows the time averaged
spatial distribution of the number of particles, which feel a pressure
$P_i > 3000\,g~cm~sec^{-2}$. These results compare directly to the
experiment by Rolf and Rothkegel\,\cite{Rothkegel} (see above). The
maxima of the spatial distribution are located at the bottom of the
cylinder near the wall which coincides with the results of the
experiment.

\begin{figure}[p]
\centerline{\psfig{figure=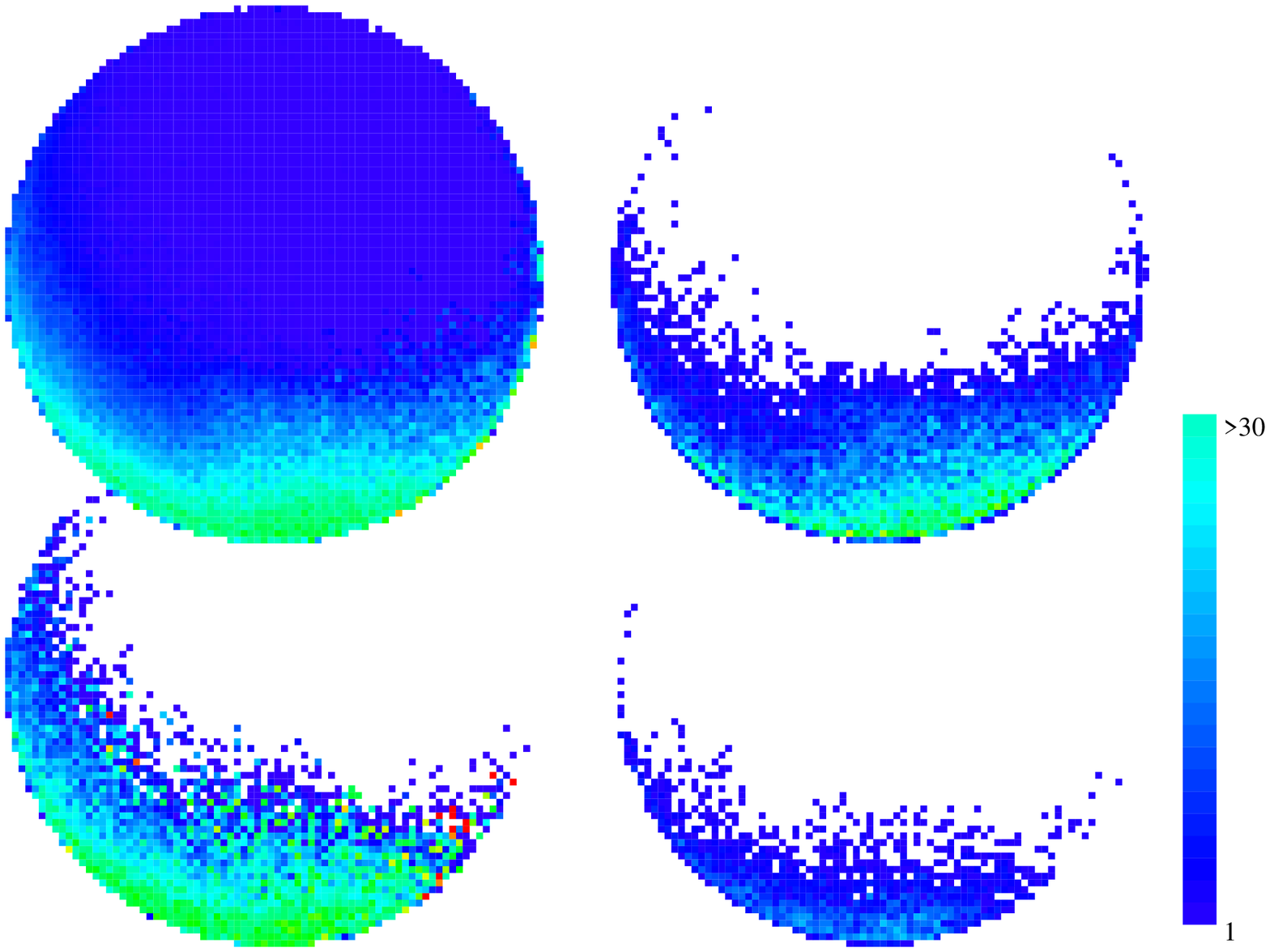,width=12cm}}
\caption{Equivalent figure to fig.~\ref{dist2} for 
  $\Omega_{III} = 19 Hz$. For explanation see fig.~\ref{dist2}.}
\label{dist19}
\end{figure}

Figures \ref{dist10} and \ref{dist19} are equivalent to
fig.~\ref{dist2} where the angular velocity of the mill is
$\Omega_{II} = 10\, Hz$ and $\Omega_{III} = 19\, Hz$, respectively.
The color scaling is the same for all left side figures and the color
code for the right figures is given by the color bar.

\begin{figure}[p]
\centerline{\psfig{figure=dist10.ps,width=12cm}}
\caption{Equivalent figure to fig.~\ref{dist2} for 
  $\Omega_{II} = 10 Hz$. For explanation see fig.~\ref{dist2}.}
\label{dist10}
\end{figure}

From these figures we conclude the surprising result that direct
impacts of particles with high relative velocities at or close to the
surface do {\em not} lead to high stresses. Therefore they are almost
neglectable for the comminution process. Regions of high pressure can
be found near the walls deep inside the material. Particles which do
not belong to a force chain feel everywhere in the material a low
pressure, only particle which are a member of a force chain feel
larger stresses, i.e. only particles which belong to a force chain
have a high probability to break. Fig.~\ref{dist19} shows that the
mechanism of force chains is of low efficiency for rather high rotation
velocities. The pressure is much lower than in the other
figures~\ref{dist2} and \ref{dist10}, which leads to a diminution in
the breaking rate as observed experimentally\,\cite{AllgBuch}.

For $\Omega = \Omega_{III}$ (fig.~\ref{dist19}) the rates resulting
from impacts of particles at the surface overcomes the rates resulting
from the force chains. Nevertheless the rates are clearly lower than
for slower spinning mill.

\section{Conclusion}
Using two-dimensional molecular dynamics simulation of spherical
granular particles we investigated the motion of granular material
moving in a ball mill which spins with different angular velocity.
The numerical results which agree with experimental data by Rolf and
Rothkegel\,\cite{Rothkegel} lead to an explanation for the
experimentally observed spatial distribution of pressure in such a
ball mill. A network of force chains which permanently changes its
structure has been observed in the rotating cylinder. Detailed and
separate analysis of the statistical properties of these chains and of
the pressure acting on particles which either {\em do} belong to a
chain, or which {\em do not} belong to a chain lead to the
conclusion that force chains play an important role for comminution
processes in granular materials. Force chains lead to concentration of
the dynamic and static load of a large amounts of grains in the bulk
of the material to few grains close to the bottom of the container.
These particles finally feel very large pressure. The described
mechanism of the action of force chains is able to interpretate
experimental data while other explanations fail\,\cite{Rothkegel}.


\section*{References}
\begin{thebibliography}{99}
  
\bibitem{Rothkegel} Rothkegel, B.: \newblock {\em \"Ortliche
    Verteilung der Sto{\ss}energien $\ge$ 22 mJ und die zugeordneten
    Bewegungszust\"ande von Modellmahlk\"orpern in einer
    Modellkugelm\"uhle,} \newblock Dissertation TU-Berlin,
  Verfahrenstechnik, 1992.
  
\bibitem{AllgBuch} Bernotat, S. and K. Sch\"onert: Size reduction (pp.
  5-1). In: {\em Ullmann's Encyclopedia of Industrial Chemistry}, VCH
  Verlag, Weinheim (1988).
  
\bibitem{FCDuran} Duran J., T. Mazozi, S. Luding, E. Cl\'ement, and J.
  Rajchenbach: Discontinuous decompaction of a falling sandpile, {\em
    Rhys. Rev. E} {\bf 53}, 1923, 1996.
  
\bibitem{FCBehringer} Behringer, R.: The Dynamics of Flowing Sand,
  {\em Nonlinear Science Today} {\bf 3}, 1, 1993.
  
\bibitem{FCDantu} Dantu, P.: Contribution \`a l'Etude M\'echanique et
  G\'eom\'etrique des Milieux Pulv\'erulents, {\em Ann. Ponts
    Chuasses} {\bf 4}, 144, 1967.
  
\bibitem{FCLiu} Liu, C. H., S. R. Nagel, D. A. Schecter, S. N.
  Coppersmith, S. Majumdar, O. Narayan, and T. A. Witten: Force
  Fluctuations in Bead Packs, {\em Science} {\bf 269}, 513, 1995.
  
\bibitem{FCWolf} Wolf, D. E.: Modelling and Computer Simulation of
  Granular Media, in {\em Computational Physics} , edited by K. H.
  Hoffmann and M. Schreiber, Springer, Heidelberg (1996).
  
\bibitem{FCEP} Esipov, S. E. and T. P\"oschel: The Granular Phase
  Diagram, {\em J. Stat. Phys., in press}.
  
\bibitem{Rajchenbach} Rajchenbach, J.: Flow in Powders: From Discrete
  Avalanches to Continuous Regime, {\em Phys.~Rev.~Lett.} {\bf 65},
  2221, 1990.
  
\bibitem{PBPRL} P\"{o}schel, T. and V. Buchholtz: \newblock Static
  friction phenomena in granular materials: Coulomb law versus
  particle geometry, \newblock {\em Phys. Rev. Lett.} {\bf 71}, 3963,
  1993.
  
\bibitem{BPDrei} P\"{o}schel, T. and V. Buchholtz: \newblock Molecular
  dynamics of arbitrarily shaped granular particles, \newblock {\em
    J.~Phys.~I France} {\bf 5}, 1431, 1995.

\bibitem{BPJSP} Buchholtz, V. and T. P\"oschel: \newblock Avalanche
  statistic of sand heaps, \newblock {\em J.~Stat.~Phys.} {\bf 84},
  1373, 1996.
  
\bibitem{CS} Cundall, P. A. and O.~D.~L.~Strack: A Discrete Numerical
  Model for Granular Assemblies, {\em G\'eotechnique} {\bf 29}, 47,
  1979.
  
\bibitem{HW} Haff, P. K. and B. T. Werner: Computer Simulation of the
  Mechanical Sorting of Grains, {\em Powder Technology} {\bf 48}, 239,
  1986.
  
\bibitem{WolfForce} Sch\"afer, J., S.~Dippel, and D.~E.~Wolf: Force
  schemes in simulations of granular materials. {\em J.~de~Phys.~I}
  {\bf 6}, 5, 1996.
  
\bibitem{AllenTildesley} Allen, M. P. and D. J. Tildesley: {\em
    Computer Simulations of Liquids}, Clarendon Press, Oxford (1987).
  
\bibitem{Weibull} Weibull, W.: The phenomenon of rupture in solids
  In\-gen\-i\"or\-ve\-tensskaps\-akademiens Handlingar, 149 (1938).

\end{thebibliography}
\end{document}